\documentclass[aps,pra,floatfix,twocolumn,groupedaddress,superscriptaddress,nofootinbib,notitlepage,amsmath,amssymb,]{revtex4-1}
\usepackage{times,graphics,graphicx,bm,bbm,bbold,amsfonts,mathtools,dsfont,float,hyperref}
\hypersetup{colorlinks,linkcolor={blue},citecolor={blue},urlcolor= {blue}}
\usepackage[normalem]{ulem}

\newcommand\cysout{\bgroup\markoverwith{\textcolor{cyan}{\rule[0.5ex]{2pt}{1.2pt}}}\ULon}
\newcommand\osout{\bgroup\markoverwith{\textcolor{orange}{\rule[0.5ex]{2pt}{1.2pt}}}\ULon}

\usepackage{amsthm}

\DeclarePairedDelimiterX\braket[2]{\langle}{\rangle}{#1 \delimsize\vert #2}

\def\dbar{{\mathchar'26\mkern-12mu \mathrm{d}}}

\let\oldsqrt\sqrt
\def\sqrt{\mathpalette\DHLhksqrt}
\def\DHLhksqrt#1#2{%
\setbox0=\hbox{$#1\oldsqrt{#2\,}$}\dimen0=\ht0
\advance\dimen0-0.2\ht0
\setbox2=\hbox{\vrule height\ht0 depth -\dimen0}%
{\box0\lower0.4pt\box2}}

\DeclareFontFamily{OT1}{pzc}{}
\DeclareFontShape{OT1}{pzc}{m}{it}%
              {<-> s * [1.25] pzcmi7t}{}
\DeclareMathAlphabet{\mathpzc}{OT1}{pzc}%
                                 {m}{it}

\begin{document}

\title{Impact of nonideal cycles on the efficiency of quantum heat engines}

\author{M. Ramezani}
\affiliation{Department of Physics, Sharif University of Technology, Tehran 14588, Iran}
\affiliation{School of Physics, Institute for Research in Fundamental Sciences (IPM), Tehran 19395, Iran}

\author{S. Marcantoni}
\affiliation{Department of Physics, University of Trieste, I-34151 Trieste, Italy}
\affiliation{National Institute for Nuclear Physics (INFN), Trieste Section, I-34151 Trieste, Italy}

\author{F. Benatti}
\affiliation{Department of Physics, University of Trieste, I-34151 Trieste, Italy}
\affiliation{National Institute for Nuclear Physics (INFN), Trieste Section, I-34151 Trieste, Italy}

\author{R. Floreanini}
\affiliation{National Institute for Nuclear Physics (INFN), Trieste Section, I-34151 Trieste, Italy}

\author{F. Petiziol}
\affiliation{Department of Mathematical, Physical, and Computer Sciences, University of Parma, 43124 Parma, Italy}
\affiliation{National Institute for Nuclear Physics (INFN), Milan Bicocca Section, Parma Group, 43124 Parma, Italy}

\author{A. T. Rezakhani}
\affiliation{Department of Physics, Sharif University of Technology, Tehran 14588, Iran}

\author{M. Golshani}
\affiliation{Department of Physics, Sharif University of Technology, Tehran 14588, Iran}
\affiliation{School of Physics, Institute for Research in Fundamental Sciences (IPM), Tehran 19395, Iran}

\date{\today}

\begin{abstract}
Given a quantum heat engine that operates in a cycle that reaches maximal efficiency for a time-dependent Hamiltonian $H(\tau)$ of the working substance, with overall controllable driving $H(\tau)=g(\tau)\,H$, we study the deviation of the efficiency from the optimal value due to a generic time-independent perturbation in the Hamiltonian. We show that for a working substance consisting of two two-level systems, by suitably tuning the interaction, the deviation can be suppressed up to the third order in the perturbation parameter---and thus almost retaining the optimality of the engine.
\end{abstract}

\maketitle

\section{Introduction}
\label{sec:intr}

The notion of a ``quantum heat engine" (QHE) was first proposed by Scovil and Schultz-DuBois \cite{1959-Scovil} in their approach to the working of a laser, whereby they interpreted a three-level atom as the working substance of the QHE and obtained the Carnot bound for the efficiency of their engine. Later, open quantum systems were considered as models of QHEs, where notions of heat and work needed careful definitions \cite{1979-Alicki,1984-Kosloff,2016-Alipour}. Investigation of QHEs has been recently reinvigorated mainly due to experimental advances in manipulating few-atom systems---see, e.g., Ref. \cite{2016-Rosnagel}.
	
Studying how quantum mechanics affects the efficiency of heat engines at microscopic scale is a central subject in quantum thermodynamics. Part of theoretical studies has focused on analyzing the efficiency of QHEs operating in the quasi-static limit \cite{2000-Bender, 2006-Quan, 2007-Quan, 2009-Quan, 2012-Brunner, 2013-Huang, 2015-Altintas}, using microscopic systems such as a two-level or multi-level atom (or, respectively, ``qubit" and ``qudits," following the jargon of quantum information science) as working substances. However, in the quasi-static limit, any heat engine, classical or quantum, even an ideal Carnot engine, independently of their efficiency, needs---by definition---an infinite (considerably long) time to operate, thence providing (almost) null power. Thus, approaching the Carnot efficiency at nonzero power has become a relevant issue per se and QHEs operating in finite times have been extensively investigated recently \cite{1992-Kosloff,1996-Kosloff,Feldmann,2011-Skrzypczyk,2011-Benenti,2012-Wang,2013-Gelbwaser,2013-Allahverdyan,2014-Kosloff,2015a-Proesmans,2015b-Polettini,2015c-Brandner,2016-Campisi}. For example, in Ref. \cite{2016-Campisi} a QHE at nonzero power and the working substance at a second-order phase transition has been considered which is argued to reach the Carnot efficiency. 
	
Another relevant subject in the study of QHEs regards how the efficiency and power of QHEs are affected by quantum effects. For example, Ref. \cite{2012-Abe} reports that superpositions of quantum states can enhance the efficiency of a QHE; Ref. \cite{2013-Goswami} reports effects of quantum coherence on the power output; Refs. \cite{2011-DeLiberato, 2012-Zagoskin, 2014-Rosnagel} consider the possibility of substituting thermal baths with the so-called squeezed baths; and Ref. \cite{2015-Uzdin} reports a bound on the power of a QHE which enables to identify the presence of quantum coherence during cycles. It is also interesting to note that although classical extractable work from a classical heat engine scales linearly with the number of cycles, the situation can differ for QHEs \cite{2017-Watanabe}.	

In this paper we analyze how a QHE with optimal efficiency may be affected by perturbations and how or when one may remedy the decrease of its efficiency and revert it to the optimal value. In particular, we consider a QHE in quasi-static thermal equilibrium with respect to a driving Hamiltonian $H(\tau)$, operating in a cycle consisting of two \textit{isothermal} and two \textit{isentropic} processes. The cycle and the driving are initially chosen such that the QHE attains the (optimal) Carnot efficiency; whereas, a small perturbation $H(\tau)\mapsto H^{[\lambda]}(\tau)=H(\tau)+\lambda V$ diminishes this optimality. We provide an explicit expression for the deviation from the maximal efficiency up to the first nonvanishing order in the perturbation parameter $\lambda$, indicating that this deviation increases by decreasing the temperature of the cold bath. Moreover, although in general the dominant terms in the deviation are of the first-order, we discuss instances of working substances consisting of one and two qubits where the second-order deviation becomes dominant. In addition, we show that the latter second-order deviations can be made vanish by suitably tuning the perturbation in the case of the interacting qubits.

This paper is organized as follows. In Sec. \ref{sec:Id}, we introduce our model of the QHE for a generic working substance, which reaches the Carnot efficiency. In Sec. \ref{sec:generic}, the presence of a small perturbation in the system Hamiltonian is explored in terms of perturbative corrections to the thermodynamic variables and loss of efficiency. In Sec. \ref{sec:twolevel}, we apply our general results to the case of two different working substances, namely for a single-qubit system and also for two interacting qubits, and analyze the possibility of countering the efficiency degradation.

\begin{figure*}[tp]
	\includegraphics[width=0.9\linewidth]{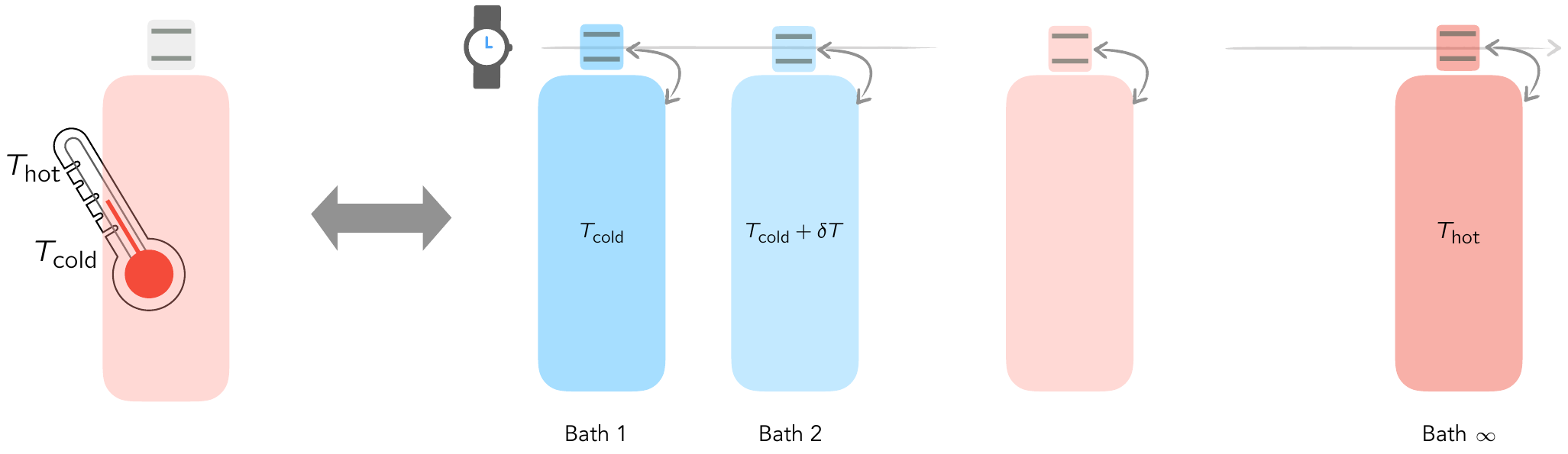} 
	\caption{The system in thermal equilibrium with a \textit{controllable} heat bath with time-dependent temperature. This scenario  can be alternatively described as the system in thermal equilibrium with infinite heat baths each with a different, constant temperature. A detailed analysis of such a control and its necessary physical resources is beyond the scope of this paper---see, e.g., Ref. \cite{Callen}.}
	\label{fig-new}
\end{figure*}

\section{Optimal-efficiency cycle}
\label{sec:Id}

We consider an optimal cycle consisting of an isothermal expansion at temperature $T_{\mathrm{hot}}$, followed by an isentropic expansion from $T_{\mathrm{hot}}$ to $T_{\mathrm{cold}}$ ($< T_{\mathrm{hot}}$), an isothermal compression at temperature $T_{\mathrm{cold}}$, and concluded by another isentropic compression from $T_{\mathrm{cold}}$ to $T_{\mathrm{hot}}$. The cycle is performed by a QHE comprised of a quantum system described by a driving Hamiltonian $H(\tau)$ with discrete, positive spectrum, of the form 
\begin{equation}
\label{idH}
H(\tau)=g(\tau)H=g(\tau)\textstyle{\sum_{j}} \epsilon_{j}\vert\epsilon_j\rangle\langle\epsilon_j\vert,
\end{equation}
where $\epsilon_j\geqslant 0$ are the eigenvalues of $H$ and $g(\tau)\geqslant 0$ is a driving parameter which we assume is adjustable at will. 

\begin{figure}[bp]
	\includegraphics[width=0.7\linewidth]{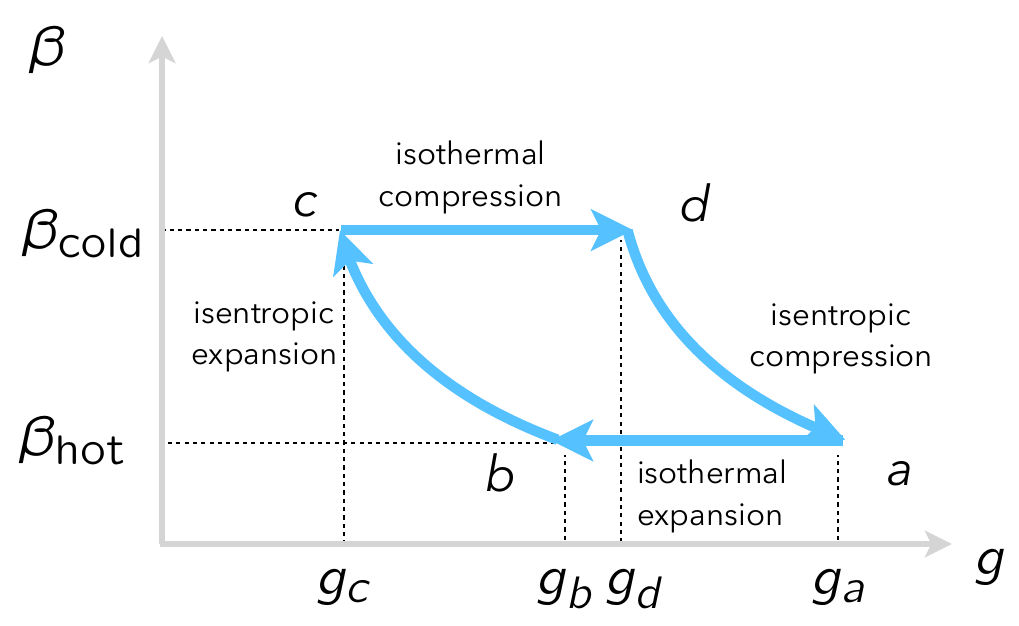} 
	\caption{The ($g,\beta$) diagram for the optimal cycle.}
	\label{FIG--Diagram}
\end{figure}

In particular, we assume such external control that all four processes can be performed ``\textit{quasi-statically};" that is, at each instant of time along the processes, the system can be considered in thermal equilibrium with an appropriate heat bath at some (time-dependent) controllable inverse temperature $\beta(\tau)=[k_{B}T(\tau)]^{-1}$ (specified later). Alternatively, this means there are infinite heat baths with which the system is in thermal equilibrium in different times---Fig. \ref{fig-new}. Thus, the state of the working substance quantum system during the cycle has always the Gibbs form,
\begin{equation}
\label{Gibbs}
\varrho(\tau)=\big[1/Z(\tau)\big]\,\mathrm{e}^{-\beta(\tau) H(\tau)},
\end{equation} 
where $Z(\tau)=\mathrm{Tr}[\mathrm{e}^{-\beta(\tau) H(\tau)}]$. We set the Boltzmann constant $k_{B}\equiv 1$ throughout the paper.

More specifically, in the $(g,\beta)$ plane, the cycle connects the initial thermodynamic state $(g_a,\beta_{\mathrm{hot}})$ to itself passing in succession through the states $(g_b,\beta_{\mathrm{hot}})$, $(g_c,\beta_{\mathrm{cold}})$, and $(g_d,\beta_{\mathrm{cold}})$, where $g_{a,b,c,d}$ are the values of the driving parameter $g(\tau)$ at the end points of the different processes as shown in Fig. \ref{FIG--Diagram}. We denote by $\varrho_{a,b,c,d}$ the thermal (Gibbs) states at the end points of the branches of the cycle, the latter ones being operationally constructed as follows. Details of thermodynamic relations used below can be found in appendix \ref{sec:H-EX}.

\subsubsection{Isothermal expansion: $a\rightarrow b$}

In the isothermal processes the system exchanges heat with a thermal bath at temperature $T_{\mathrm{hot}}$, changing its entropy. During this branch, the temperature is kept fixed at $T=T_{\mathrm{hot}}$ and $g(\tau)$ varies from an initial value $g_a$ to a final value $g_b$ such that the process corresponds to an expansion and that, at each instant of time $\tau$ during the process, the state of the system is described as in Eq. \eqref{Gibbs} with $\beta(\tau)$ fixed to $\beta_{\mathrm{hot}}$,
\begin{equation}
\label{Gibbsab1}
\varrho_{a,b}=(1/Z_{a,b})\,\mathrm{e}^{-\beta_{\mathrm{hot}}\,g_{a,b} H},
\end{equation}
where $Z_{a,b}=\sum_i\mathrm{e}^{-\beta_{\mathrm{hot}}\,g_{a,b}\epsilon_i}$. One can see that in the isothermal process $a\rightarrow b$, the work and the heat exchange of the system are given by (appendix \ref{sec:H-EX})   
\begin{align}
\label{Work_Iso}
\mathbbmss{W}_{ab}&=\mathbbmss{F}(b)-\mathbbmss{F}(a),\\
\mathbbmss{Q}_{ab}&=(1/\beta_{\mathrm{hot}})[\mathbbmss{S}(b)-\mathbbmss{S}(a)],  
\end{align}
where $\mathbbmss{F}(a)$ and $\mathbbmss{S}(a)$ (similarly $\mathbbmss{F}(b)$ and $\mathbbmss{S}(b)$) are the free energies and von Neumann entropies relative to the state $\varrho_a$ ($\varrho_b$) at the end and beginning of the process. Moreover, we have 
\begin{equation}
\label{procabF}
\mathbbmss{F}(a)=-(1/\beta_{\mathrm{hot}})\log Z_a,\quad \mathbbmss{F}(b)=-(1/\beta_{\mathrm{hot}})\log Z_b.
\end{equation}
Choosing $g_b\leqslant g_a$ ensures $Z_b\geqslant Z_a$, and hence $\mathbbmss{W}_{ab}\leqslant 0$ is the work performed \textit{by} the system, and the isothermal process $a\rightarrow b$ is an \textit{expansion}. If we lighten the notation and denote $\varrho(\tau)$ simply by the corresponding $\varrho_{g}$, the von Neumann entropy of the time-varying state \eqref{Gibbs} reads
\begin{align}
\mathbbmss{S}(g)&=-\mathrm{Tr}[\varrho_g\log\varrho_g]\nonumber\\
&=\log Z_g+\frac{\beta g}{Z_g} \textstyle{\sum_j} \epsilon_j\,\mathrm{e}^{-\beta g\epsilon_j}, \label{iso-heat1}
\end{align}
with $Z_g=\sum_j\mathrm{e}^{-\beta g\epsilon_j}$. Note that 
 \begin{equation}
\label{iso-heat2}
\partial_g \mathbbmss{S}(g)=-\beta^{2}g\big(\mathrm{Tr}[\varrho_g\,H^2]-\big(\mathrm{Tr}[\varrho_{g}\,H]\big)^2\big)\leqslant 0.
\end{equation}
Using the expression  
\begin{equation}
\label{QQ}
\mathbbmss{Q}=\textstyle{\int_{\mathrm{i}}^{\mathrm{f}}} (1/\beta)\, \mathrm{d}\mathbbmss{S},
\end{equation}
where the integration is performed over the path in the parameter space that identifies the branch connecting its initial and final states ``$\mathrm{i}$" and ``$\mathrm{f}$," Eq. (\ref{iso-heat2}) yields $\mathbbmss{Q}_{ab}\geqslant 0$; hence, the heat is absorbed by the system from the hot bath.

\subsubsection{
Isentropic expansion: $b\rightarrow c$}

This branch of the cycle is implemented by employing a \textit{controllable} heat bath as described in Fig. \ref{fig-new} and manipulating the system in a particular fashion. Specifically, we vary $T(\tau)$ quasi-statically from $T_{\mathrm{hot}}$ to $T_{\mathrm{cold}}$ ($\leqslant T_{\mathrm{hot}}$) and concurrently change the driving parameter $g(\tau)$ of the system from the initial value $g_b$ to the final value $g_c$ such that during this whole quasi-static process the quantum system performs work (expansion), whereas the product $\beta(\tau)\,g(\tau)=\mathrm{const.}$, and thus the system can be described by a \textit{fixed} thermal state,
\begin{align}
\label{adiab1}
\beta(\tau)\,g(\tau) &=\beta_{\mathrm{hot}}\,g_b=\beta_{\mathrm{cold}}\,g_c,\\
\varrho(\tau) &=\varrho_b=\varrho_c.
\end{align}
Thus the heat exchange during this process vanishes,\, $\dbar \mathbbmss{Q}=0$ and the process becomes ``isentropic," $\mathbbmss{S}(b)=\mathbbmss{S}(c)$---for details of the calculations see appendix \ref{sec:H-EX}.

\subsubsection{Isothermal compression: $c\rightarrow d$}

While the temperature is kept fixed at $T_{\mathrm{cold}}$, the driving $g(\tau)$ is chosen such that the system isothermally releases an amount of heat $\mathbbmss{Q}_{cd}\leqslant 0$ to the cold bath and is subjected to compression,
\begin{align}
\label{Work_Iso2}
\mathbbmss{Q}_{cd} &=(1/\beta_{\mathrm{cold}})[\mathbbmss{S}(d)-\mathbbmss{S}(c)]\leqslant 0,\\
\mathbbmss{W}_{cd} &=\mathbbmss{F}(d)-\mathbbmss{F}(c)\geqslant 0. 
\end{align}

\subsubsection{Isentropic compression: $d\rightarrow a$}

The cycle is closed by an isentropic compression where $\beta(\tau)$ is varied to increase the temperature of the system from $T_{\mathrm{cold}}$ to $T_{\mathrm{hot}}$ and the driving parameter $g(\tau)$ is varied from the initial value $g_d$ to the final value $g_a$ such that the quasi-static process is characterized by
\begin{equation}
\label{adiab2}
\beta(\tau)\,g(\tau)=\beta_{\mathrm{cold}}\,g_d=\beta_{\mathrm{hot}}\,g_a,\qquad \mathbbmss{S}(d)=\mathbbmss{S}(a).
\end{equation}

Since $\mathbbmss{Q}_{ab}\geqslant 0$, $\mathbbmss{Q}_{cd}\leqslant 0$, and $\mathbbmss{S}(b)-\mathbbmss{S}(a)=\mathbbmss{S}(c)-\mathbbmss{S}(d)$, the efficiency $\eta$ of the cycle is maximal,
\begin{align}
\eta &=1-\frac{\left|\mathbbmss{Q}_{cd}\right|}{\mathbbmss{Q}_{ab}}\nonumber\\
 & =1-\frac{\beta_{\mathrm{hot}}}{\beta_{\mathrm{cold}}}\,\frac{\mathbbmss{S}(c)-\mathbbmss{S}(d)}{\mathbbmss{S}(b)-\mathbbmss{S}(a)}\nonumber\\
& =1-\frac{T_{\mathrm{cold}}}{T_{\mathrm{hot}}}, \label{Carnoteff}	
\end{align}
which is the Carnot efficiency.

\section{Perturbed optimal cycle}
\label{sec:generic}

The QHE operates the optimal cycle designed in the previous section by varying one or both of the external parameters, namely the inverse temperature $\beta(\tau)$ and the driving $g(\tau)$, in such a way to implement the two isothermal and the two isentropic processes in Fig. \ref{FIG--Diagram}, the latter ones in closed connection with the form of the Hamiltonian, $H(\tau)=g(\tau)H$, that is by keeping $\beta(\tau)\,g(\tau)$ constant.

Suppose the QHE operates the same cycle, that is by varying $\beta(\tau)$ and $g(\tau)$ exactly as for the case of a quantum medium with Hamiltonian $H(\tau)$, however using a quantum system with Hamiltonian 
\begin{equation}
\label{perturb0}
H^{[\lambda]}(\tau)=g(\tau) H+\lambda V,
\end{equation}
where $V=V^\dag$ is a perturbation and $\lambda$ a relatively small perturbative parameter along the various processes ($|\lambda| \Vert V\Vert\ll |g(\tau)|\Vert H\Vert$, where $\Vert\,\Vert$ is the standard operator norm). The latter is again assumed to be quasi-static so that at each instant of time the system states will be of the Gibbs form,
\begin{equation}
\label{pertGibbs1}
\varrho^{[\lambda]}(\tau)=\big(1/Z^{[\lambda]}(\tau)\big)\mathrm{e}^{-\beta(\tau)(g(\tau) H+\lambda V)},
\end{equation}
where $Z^{[\lambda]}(\tau)=\mathrm{Tr}[\mathrm{e}^{-\beta(\tau)(g(\tau)H+\lambda V)}]$.

Certainly, while the isothermal branches will remain so, the two processes that were isentropic for $H(\tau)$ are no longer such for $H^{[\lambda]}(\tau)$, there will be heat exchanges $\mathbbmss{Q}_{bc}$ and $\mathbbmss{Q}_{da}$ also in the branches $b\rightarrow c$ and $d\rightarrow a$ whence the entropies at the beginning and end of them will not be equal anymore and the efficiency of the QHE will deviate from optimality. Our aim is to expand the efficiency of the perturbed QHE as a power series in the small parameter $\lambda$. More practically, we shall be content with obtaining the first nonvanishing correction, say $k $, that is we will seek an approximation to $\eta_\lambda$ of the form 
\begin{equation}
\label{effexp}
\eta(\lambda)=1-\frac{\left|\mathbbmss{Q}_{\rightarrow}(\lambda)\right|}{\mathbbmss{Q}_{\leftarrow}(\lambda)}\approx \eta_0\,-\,\lambda^{k}\eta_{k},
\end{equation}
where $\mathbbmss{Q}_{\leftarrow}(\lambda)\geqslant 0$ and $\mathbbmss{Q}_{\rightarrow}(\lambda)\leqslant 0$ are, respectively, the heat absorbed and released during the cycle with the perturbed Hamiltonian, and $\eta_{0}$ is the maximal Carnot efficiency \eqref{Carnoteff}.
Let us expand the heat exchanges $\mathbbmss{Q}_{if}(\lambda)$ from an initial thermal equilibrium state ``$\mathrm{i}$" to a final thermal equilibrium state "$\mathrm{f}$" up to their first non-vanishing term, 
\begin{equation}
\label{heatpertgen}
\mathbbmss{Q}_{\mathrm{if}}(\lambda)=\mathbbmss{Q}_{\mathrm{if}}^{(0)}+\lambda^{k}\mathbbmss{Q}^{(k )}_{\mathrm{if}}\ ,
\end{equation}
where $\mathbbmss{Q}_{\mathrm{if}}^{(0)}$ and $\mathbbmss{Q}_{\mathrm{if}}(\lambda)$ denote the heat exchanges without ($\lambda=0$) and with perturbative effects.
Along the isothermal processes one has
$\mathbbmss{Q}_{ab}^{(0)}=\mathbbmss{Q}_{ab}\geqslant 0$, $\mathbbmss{Q}_{cd}^{(0)}=\mathbbmss{Q}_{cd}\leqslant 0$ with $\mathbbmss{Q}_{ab}$ and $\mathbbmss{Q}_{cd}$ the heat exchanges during the optimal cycle without perturbation. One can thus
always assume $\lambda$ sufficiently small so that $\mathbbmss{Q}_{ab}(\lambda)\geqslant 0$ and $\mathbbmss{Q}_{cd}(\lambda)\leqslant 0$ up to that truncation of the series expansion. Thus $\mathbbmss{Q}_{ab}(\lambda)$ and $\mathbbmss{Q}_{cd}(\lambda)$ will be considered as heat absorbed and released, respectively. 

Note that the zeroth-order contribution in the isentropic branches vanishes and the $k$th-order can be assigned either to $\mathbbmss{Q}_{\rightarrow}$ or $\mathbbmss{Q}_{\leftarrow}$ depending on the case. In fact, now the branches $b\to c$ and $d\to a$ of the cycle, that are isentropic in the absence of perturbation, when $\lambda\neq0$ also allow for heat exchanges and thus contribute
to the total amount of heat absorbed and released,
\begin{align}
\label{pertheat1}
\mathbbmss{Q}_{\leftarrow}(\lambda) \approx\ & \mathbbmss{Q}_{ab}+\lambda^{k} \mathbbmss{Q}^{(k )}_{\leftarrow},\\
\label{pertheat2}
\mathbbmss{Q}_{\rightarrow}(\lambda) \approx\ & \mathbbmss{Q}_{cd}+\lambda^{k}\mathbbmss{Q}^{(k )}_{\rightarrow} .
\end{align}
Since $\mathbbmss{Q}_{\rightarrow}(\lambda)\leqslant 0$ (for sufficiently small $\lambda$), inserting $\left|\mathbbmss{Q}_{\rightarrow}(\lambda)\right|=\left|\mathbbmss{Q}_{cd}\right|-\lambda^{k}\,\mathbbmss{Q}^{(k )}_{\rightarrow}$ into Eq. \eqref{effexp} and using $\left|\mathbbmss{Q}_{cd}\right|/\mathbbmss{Q}_{ab}=\beta_{\mathrm{hot}}/\beta_{\mathrm{cold}}$, one computes
\begin{equation}
\label{etak}
\eta_{k}=-\frac{\mathbbmss{Q}^{(k )}_{\rightarrow}}{\mathbbmss{Q}_{ab}}-\frac{\beta_{\mathrm{hot}}}{\beta_{\mathrm{cold}}}\frac{\mathbbmss{Q}^{(k )}_{\leftarrow}}{\mathbbmss{Q}_{ab}}\ .
\end{equation}
As we shall see later in the examples, the above correction is always nonnegative so that the efficiency can only become less than or remain the same as the maximal one.

In order to extract an explicit expression for the deviation $\eta_{k}$, note that the heat exchanges are again given by Eq. \eqref{Work_Iso} along the isothermal branches and by Eq. (\ref{QQ}) along the branches $b\rightarrow c$ and $d\rightarrow a$. Given a generic thermal Gibbs state, from equilibrium thermodynamic relations (see appendix \ref{sec:H-EX}) we have
\begin{equation}
\label{ent}
\mathbbmss{S}=-\beta^2\partial_\beta[(1/\beta)\log Z].
\end{equation}
The perturbed partition function $Z^{[\lambda]}$ can be expanded as a power series of $\lambda$ by means of the Dyson series method---appendix \ref{sec:QHE}. One then finds that, up to the first nonvanishing order in $\lambda$, 
\begin{equation}
\label{pertZ}
Z^{[\lambda]} \approx  Z_0(\beta g)+(\beta\lambda)^{k} Z_{k}(\beta g),
\end{equation}
where the zeroth-order contribution $Z_0(\beta g)$ is given in Eq. \eqref{Gibbs} and the first nonvanishing correction $Z_{k}(\beta g)$ is reported in appendix \ref{sec:QHE} (note that both terms depend only on the product quantity $\beta g$). Inserting the expanded partition function into Eq. \eqref{ent}, up to the first nonvanishing correction in $\lambda$, the entropy can similarly be approximated by
\begin{equation}
\label{pertent1}
\mathbbmss{S}^{[\lambda]}\approx  \mathbbmss{S}_0(\beta g)+(\lambda\beta)^{k}\, \mathbbmss{S}_{k}(\beta g),
\end{equation}
where the zeroth-order term is the von Neumann entropy in the ideal case,
\begin{equation}
\label{pertent2}
\mathbbmss{S}_0(\beta g)=\log Z_0(\beta g)-\beta\partial_\beta\,\log Z_0(\beta g),
\end{equation}
and the first nonvanishing correction reads
\begin{equation}
\label{pertent3}
\mathbbmss{S}_{k} (\beta g)=(-)^{k +1}\,\beta^{2-k }\partial_\beta\left(\beta^{k -1}\,\frac{Z_{k}(\beta g)}{Z_0(\beta g)}\right),
\end{equation}
and here too both contributions depend only on the product $\beta g$---as shown in appendix \ref{sec:QHE}. Then, with the notation of Eq. \eqref{heatpertgen} and using the conditions \eqref{adiab1} and \eqref{adiab2}, one can obtain
\begin{align}
\mathbbmss{Q}^{(k )}_{ab}=&\frac{\mathbbmss{S}_{k}(\beta_{\mathrm{hot}}\,g_b) - \mathbbmss{S}_{k}(\beta_{\mathrm{hot}}\,g_a)}{\beta^{1-k}_{\mathrm{hot}}}, \label{pertheatisoth1}\\
\mathbbmss{Q}^{(k )}_{cd}=&\frac{\mathbbmss{S}_{k}(\beta_{\mathrm{cold}} \,g_d)-\mathbbmss{S}_{k}(\beta_{\mathrm{cold}}\, g_c)}{\beta^{1-k}_{\mathrm{cold}}}=
-\mathbbmss{Q}^{(k )}_{ab}\frac{\beta^{k -1}_{\mathrm{cold}}}{\beta^{k -1}_{\mathrm{hot}}}. \label{pertheatisoth2}
\end{align}

Concerning the heat exchanges during the ideally isentropic processes $b\rightarrow c$ and $d\rightarrow a$, along them the product term $\beta g$ is constant so that $\mathrm{d}\mathbbmss{S}_0(\beta g)=0$, and consequently $\mathbbmss{Q}^{(0)}_{bc}=\mathbbmss{Q}^{(0)}_{da}=0$. On the other hand, noting
\begin{equation}
\mathrm{d}\left[(\lambda\beta)^{k} \mathbbmss{S}_{k}(\beta g)\right]=k  \lambda^{k} \beta^{k -1}\, \mathbbmss{S}_{k}(\beta g)\,\mathrm{d}\beta
\end{equation}
and using Eq. (\ref{QQ}) one can obtain
\begin{align}
\label{pertheatad1}
\mathbbmss{Q}_{bc}(\lambda)\approx &\ \lambda^{k}\,\frac{k}{k -1}\left(\beta_{\mathrm{cold}}^{k -1}-\beta_{\mathrm{hot}}^{k -1}\right)\mathbbmss{S}_{k}(\beta_{\mathrm{hot}} \, g_b),
\\
\label{pertheatad2}
\mathbbmss{Q}_{da}(\lambda)\approx &\ \lambda^{k}\,\frac{k}{k -1}\left(\beta_{\mathrm{hot}}^{k -1}-\beta_{\mathrm{cold}}^{k -1}\right)\mathbbmss{S}_{k}(\beta_{\mathrm{cold}} \, g_d),
\end{align}
for $k \geqslant 2$, with a simple logarithmic limit for $k =1$.  

Although $\mathbbmss{Q}_{ab}(\lambda)\geqslant 0$ and $\mathbbmss{Q}_{cd}(\lambda)\leqslant 0$, the signs of $\mathbbmss{Q}_{bc}(\lambda)$ and $\mathbbmss{Q}_{da}(\lambda)$ are in general not fixed as they depend on the signs of $\mathbbmss{S}_{k}(\beta_{\mathrm{hot}}\, g_b)$ and $\mathbbmss{S}_{k}(\beta_{\mathrm{cold}} \,g_d)$. For instance, if $\mathbbmss{S}_{k}(\beta_{\mathrm{hot}} \, g_b)$ and $\mathbbmss{S}_{k}(\beta_{\mathrm{cold}}\, g_d)$ are both positive, then $\beta_{\mathrm{cold}}\geqslant \beta_{\mathrm{hot}}$ makes $\mathbbmss{Q}_{bc}(\lambda)\geqslant 0$ represent the absorbed heat and $\mathbbmss{Q}_{da}(\lambda)\leqslant 0$ released heat. Then, insertion of Eqs. \eqref{pertheatisoth1} -- \eqref{pertheatad2} into Eqs. \eqref{pertheat1} and \eqref{pertheat2} and next in Eq. \eqref{etak} and assuming $\gamma=\beta_{\mathrm{hot}}/\beta_{\mathrm{cold}}$ yield
\begin{equation}
\eta_{k}=\frac{\beta_{\mathrm{hot}}\,\beta_{\mathrm{cold}}^{k -1}}{\mathbbmss{S}(b)-\mathbbmss{S}(a)}\Big(u_{k}(\gamma)\mathbbmss{S}_{k}(\beta_{\mathrm{hot}}\, g_b) + v_{k}(\gamma) \mathbbmss{S}_{k}(\beta_{\mathrm{hot}}\, g_a)\Big),
\label{effpert}
\end{equation}
where $\mathbbmss{S}(a)$ and $\mathbbmss{S}(b)$ are the von Neumann entropies of the Gibbs states in Eq. \eqref{Gibbsab1} and we have introduced
\begin{align}
\label{ukg}
u_{k}(\gamma)=&1+\frac{1}{k -1}\left(\gamma^{k}-k \gamma\right),\\
\label{vkg}
v_k(\gamma)=&\gamma^{k}-\frac{1}{k -1}\left(k \gamma^{k -1}-1\right).
\end{align}
Of particular interest in the following are the first- and second-order contributions,
\begin{align}
\eta_1=&\frac{\beta_{\mathrm{hot}}}{\mathbbmss{S}(b)-\mathbbmss{S}(a)}\,\Big[(1-\gamma+\gamma\log\gamma) \mathbbmss{S}_1(\beta_{\mathrm{hot}}\,g_b) \nonumber\\
&+(\gamma-\log\gamma-1) \mathbbmss{S}_1(\beta_{\mathrm{hot}}\,g_a)\Big], \label{eta1st} \\
\eta_2=&\frac{\beta_{\mathrm{hot}}\beta_{\mathrm{cold}}}{\mathbbmss{S}(b)-\mathbbmss{S}(a)}\,(1-\gamma)^2\Big[\mathbbmss{S}_2(\beta_{\mathrm{hot}}\, g_b)+ \mathbbmss{S}_2(\beta_{\mathrm{hot}}\,g_a)\Big].
\label{eta2nd}
\end{align}
Similar expressions for the correction $\eta_{k}$ to the Carnot efficiency can be obtained for different signs of the heat contributions $\mathbbmss{Q}^{(k )}_{bc}$ and $\mathbbmss{Q}_{da}^{(k )}$.
Explicitly, one finds
\begin{equation}
\label{genform}
\eta_{k}=\beta_{\mathrm{hot}}\,\beta_{\mathrm{cold}}^{k -1}\,\frac{\mu_{k}(\gamma)\left|\mathbbmss{S}_{k}(\beta_{\mathrm{hot}} \, g_b)\right|+\nu_{k}(\gamma)\left|\mathbbmss{S}_{k}(\beta_{\mathrm{hot}}\, g_a)\right|}{\mathbbmss{S}(b)-\mathbbmss{S}(a)},
\end{equation}
where
\begin{align}
	\mu_{k}(\gamma)&=\left\{
	\begin{array}{ll}
		u_{k}(\gamma) &\quad\hbox{if}\quad \mathbbmss{S}_{k}(\beta_{\mathrm{hot}}\, g_b)>0,\\
		v_{k}(\gamma) &\quad\hbox{if}\quad \mathbbmss{S}_{k}(\beta_{\mathrm{hot}}\, g_b)<0,
	\end{array}
	\right.\\
	\nu_{k}(\gamma)&=\left\{
	\begin{array}{ll}
		v_{k}(\gamma) &\quad\hbox{if}\quad  \mathbbmss{S}_{k}(\beta_{\mathrm{hot}}\, g_a)>0,\\
		u_{k}(\gamma) &\quad\hbox{if}\quad \mathbbmss{S}_{k}(\beta_{\mathrm{hot}}\, g_a)<0.
	\end{array}\right.
\end{align} 

It appears that the deviation from the Carnot efficiency in Eq. \eqref{effexp} is always negative; indeed, the expression of $\eta_{k}$ is always positive as one can ascertain by taking the derivative of $u_{k}(\gamma)$ and $v_{k}(\gamma)$ with respect to $0\leqslant \gamma\leqslant 1$. It is straightforward to verify that both these derivatives are nonpositive for $k\geqslant 1$ so that $u_{k}(\gamma)$ monotonically decreases from $u_{k}(0)=1$ to $u_{k}(1)=0$ and $v_{k}(\gamma)$ from $v_{k}(0)=1/(k -1)$ to $v_{k}(1)=0$, both resulting always nonnegative.

\section{Applications}
\label{sec:twolevel}

We now explicitly apply the above scenarios to the cases of QHEs whose working substances are, respectively, a single-qubit and a two-qubit system. We show that, unlike in the single-qubit case, in the two-qubit case the interaction can be used to suppress the deviation from the ideal Carnot efficiency.

\subsection{One-qubit QHE}

According to the thermodynamic setting developed in the previous sections, we consider a QHE operating the optimal cycle described there acting upon a quantum medium described by a generic two-level traceless driven Hamiltonian 
\begin{equation}
\label{1qHam}
H(\tau) = \bm{g}(\tau)\cdot \bm{\sigma},
\end{equation}
where $\bm{g}(\tau)$ denotes the vector of the Hamiltonian couplings corresponding to the Pauli matrices $\bm{\sigma}=(\sigma_1, \sigma_2, \sigma_3)$. The cycle is assumed to consist of processes that are performed quasi-statically such that the system is at each instant of time in an instantaneous equilibrium thermal state 
\begin{equation}
\label{eq:thermalstate}
\varrho(\tau) = \frac{\mathbbmss{I}}{2} - \frac{\tanh(\beta(\tau)g(\tau))}{2g(\tau)}\,  \bm{g}(\tau)\cdot\bm{\sigma},
\end{equation}
with possibly time-dependent inverse temperature $\beta(\tau)$, 
where $g(\tau)=\Vert \bm{g}(\tau)\Vert$ ($\Vert \,\Vert$ here is the standard vector norm).

For simplicity, we set 
\begin{equation}
x\equiv\beta(\tau)g(\tau),
\end{equation}
then, from Eq. \eqref{ent}, the von Neumann entropy reads
\begin{equation}
\mathbbmss{S}(x) =   -x\,\tanh x +\log( 2 \cosh x).
\label{eq:tdpot}
\end{equation}

Let now the Hamiltonian of the system be perturbed by  $\lambda V$, where $\lambda>0$ is small parameter. More specifically, let us suppose that the QHE operates the optimal cycle that reaches the Carnot efficiency for a driving Hamiltonian $H(\tau)=g(\tau) \sigma_3$, whereas the actual Hamiltonian of the system is
\begin{equation}
\label{1qHampert0}
H^{[\lambda]}(\tau) =  g(\tau)\,\sigma_3 +\lambda V.
\end{equation}
The perturbation $V$ can in general be expressed as $V = \bm{v}\cdot \bm{\sigma}$ with $\bm{v}$ a real vector, thus
\begin{equation}
 \label{1qHampert1}
 H^{[\lambda]}(\tau)=\bm{g}^{[\lambda]}(\tau)\cdot\bm{\sigma},
 \end{equation}
where $ \bm{g}^{[\lambda]}(\tau)=(\lambda v_1,\lambda v_2,g(\tau)+\lambda v_3)$. By means of Eqs. \eqref{pertpart0} -- \eqref{pertpart2}, one expands the partition function as
\begin{equation}
Z^{[\lambda]}(x)=Z_0(x)+\lambda\beta(\tau)\,Z_1(x)\,+\,(\lambda\beta(\tau))^2\,Z_2(x),
\end{equation}
where $Z_0(x)=2\cosh x$ and  
\begin{align}
\label{1qpartpert1}
Z_1(x)=&2v_3\,\sinh x,\\
\label{1qpartpert2}
Z_2(x)=&v_3^2 \cosh x+\frac{\sinh x}{x}(v_1^2+v_2^2).
\end{align}
One can derive (from Eq. \eqref{appBent}) the perturbation expansion for the entropy up to $O(\lambda^3)$,
\begin{equation}
\mathbbmss{S}^{[\lambda]}(\tau) = \mathbbmss{S}_0(x) + (\lambda\beta(\tau)) \mathbbmss{S}_1(x) +(\lambda\beta(\tau))^{2} \,\mathbbmss{S}_2(x),
\end{equation}
where $\mathbbmss{S}_0(x)$ is as in Eq. \eqref{eq:tdpot}, whereas
\begin{align}
\label{subeq3}
\mathbbmss{S}_1(x)=& - v_3 \,\frac{x}{\cosh^2x},\\
\label{eq:entr2} 
\mathbbmss{S}_2(x)=& -\frac{v_3^2}{2}-\frac{v_1^2+v_2^2}{2\cosh^2 x}\ .
\end{align}
Note that, if $v_3=0$, then the first nonvanishing contribution is $\mathbbmss{S}_2(x)$, which is never positive, and that both $\mathbbmss{S}_{1,2}(x)$ are constant along the isentropic processes $b\to c$ and $d\to a$. We now discuss two cases, namely $v_3\neq 0$ so that the first nonvanishing correction to the entropy is the first-order one, $\mathbbmss{S}_1(x)$ and the case $v_3=0$ when the first contributing correction is  the second order one, $\mathbbmss{S}_2(x)$. 

\subsubsection{Case $v_3\ne 0$}

The sign of $v_3\neq 0$ is opposite to that of the first-order correction $\mathbbmss{S}_1(x)$. In particular, if $v_3 <0$, then the system releases heat during the process $d\to a$ and absorbs heat in the process $b\to c$. Then, using Eq. \eqref{eta1st} one finds 
\begin{align}
\eta_1 =& \frac{\beta_{\mathrm{hot}}}{\mathbbmss{S}(\beta_{\mathrm{hot}}g_b)- \mathbbmss{S}(\beta_{\mathrm{hot}}\,g_a)}\Big[(1-\gamma+\gamma \log \gamma) \mathbbmss{S}_1(\beta_{\mathrm{hot}}g_b)\nonumber\\ 
\label{1qeta1}
&+ (\gamma-\log \gamma-1)\, \mathbbmss{S}_1(\beta_{\mathrm{hot}}\,g_a)\Big].
\end{align}
From Eq. \eqref{subeq3} one sees that the first-order correction to the ideal Carnot efficiency can vanish only if $v_3=0$.

\subsubsection{Case $v_3 = 0$}

The first-order correction to the efficiency vanishes and the second-order contribution reads
\begin{equation}
\label{1qent2}
\mathbbmss{S}_2(x)= -\frac{v_1^2+v_2^2}{2\cosh^2 x}\ .
\end{equation}
Since it is always negative and $\beta_{\mathrm{cold}}\geqslant\beta_{\mathrm{hot}}$, the system absorbs heat in the process $d\to a$ and releases it in the process $b\to c$; then, one obtains
\begin{equation}
\label{1qeta2}
\eta_2 = \beta_{\mathrm{hot}} \beta_{\mathrm{cold}} (1-\gamma)^2\frac{\left|\mathbbmss{S}_2(\beta_{\mathrm{hot}}g_b)\right|+\left|\mathbbmss{S}_2(\beta_{\mathrm{hot}}g_a)\right|}{\mathbbmss{S}(\beta_{\mathrm{hot}}g_b)- \mathbbmss{S}(\beta_{\mathrm{hot}}g_a)}.
\end{equation}
One thus sees that, if the first-order correction $\eta_1$ to the Carnot efficiency vanishes, $v_3=0$, the second order correction never does for $V\neq 0$, that is when either $v_{1,2}\neq 0$.

\subsection{Two-qubit QHE}

In this section, we consider a QHE whose working substance is a pair of qubit systems and which operates an optimal cycle as the one discussed in Sec. \ref{sec:Id}, which is ideal with maximal Carnot efficiency when there are no interactions between the two systems and their Hamiltonian is of the form
\begin{equation}
\label{2qfree}
H(\tau)=g(\tau)\big(\sigma_3\otimes \mathbbmss{I}+\mathbbmss{I}\otimes\sigma_3\big),
\end{equation}
where $g(\tau)\geqslant 0$ is the external driving parameter.
We discuss how switching on a perturbative interaction affects the efficiency and show that, unlike in the case of one-qubit system, by suitably tuning the interaction one can still reach a maximal Carnot efficiency up to the third order in the perturbation parameter. 

\begin{figure}[tp] 
 \includegraphics[width=0.7\linewidth]{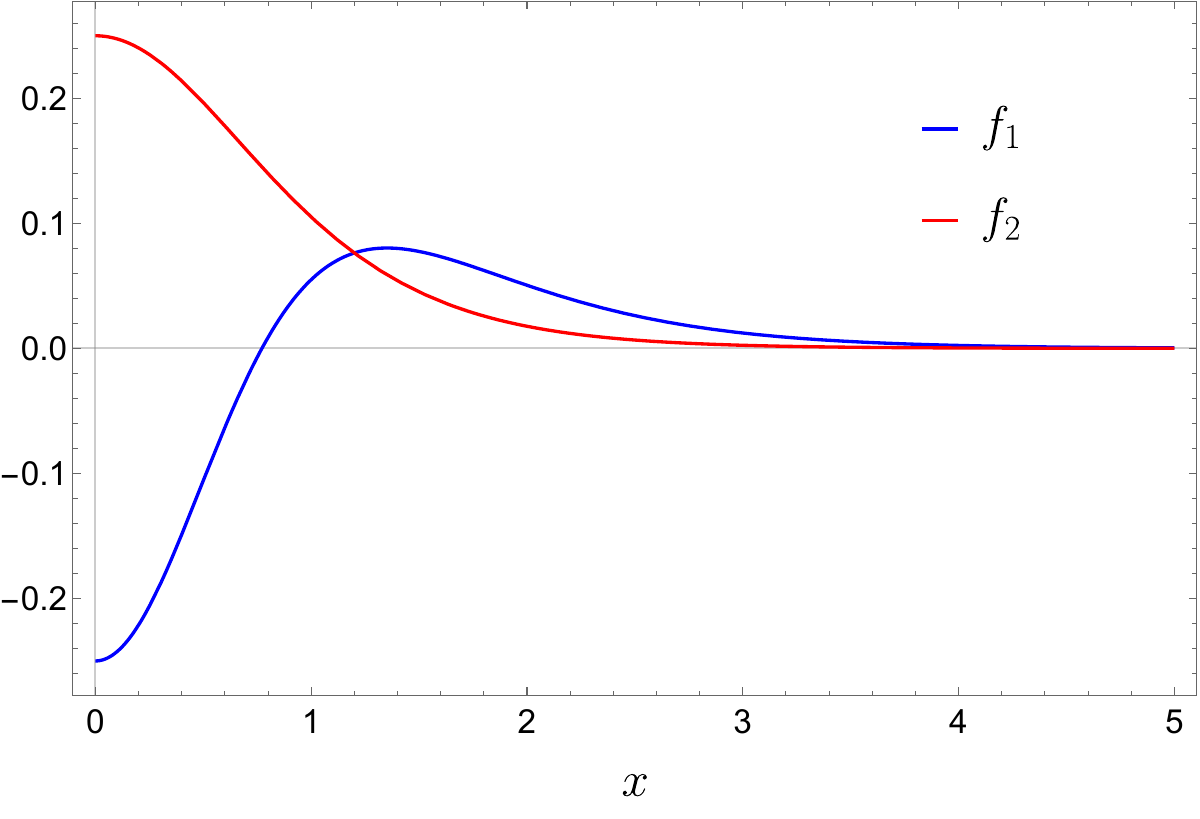} 
 \caption{$f_1(x)$ and $f_2(x)$ vs. $x$---Eqs. (\ref{2qpertent3}) and (\ref{2qpertent4}).} 
 \label{FIG2} 
\end{figure}  
Specifically, we consider a Heisenberg perturbation, namely a working substance with the Hamiltonian
\begin{align}
\nonumber
H^{[\lambda]}(\tau)=&H(\tau)\,+\, \lambda\, \Big(J_{1}(\sigma_{+}\otimes\sigma_{-}+\sigma_{-}\otimes\sigma_{+})\\
\label{2qpertham}
&+ J_{2}(\sigma_{+}\otimes\sigma_{+}+\sigma_{-}\otimes\sigma_{-})\Big) ,
\end{align}
where $\sigma_\pm=(\sigma_1\pm i\sigma_2)/2$. The Dyson expansion---appendix \ref{sec:QHE}---gives a vanishing first-order correction, $Z_1(x)=0$,
to the unperturbed partition function (see Eqs. \eqref{pertpart0} -- \eqref{pertpart2}) $Z_0(x)=4\cosh^2x$, but 
\begin{equation}
\label{2qpertpart}
Z_2(x)=J_{1}^{2}\,+\,J_2^2\frac{\sinh x\cosh x}{x}.
\end{equation}
Then, from Eq. \eqref{appBent} it follows that
\begin{eqnarray}
\label{2qpertent1}
\mathbbmss{S}_0(x)&=&\ln(4\cosh^2x)-2x\tanh x,\\
\label{2qpertent2}
\mathbbmss{S}_2(x)&=&J_{1}^{2}\,f_1(x)\,-\,J_{2}^{2}\,f_2(x),
\end{eqnarray}
where
\begin{align}
\label{2qpertent3}
f_1(x) =& \frac{2x\tanh x-1}{4\cosh^2x},\\
f_2(x) =& \frac{1}{4\cosh^2x}. \label{2qpertent4}
\end{align}
By inserting the above expression of $\mathbbmss{S}_2(x)$ into that of the deviation $\eta_2$ in Eq. \eqref{1qeta2}, one can always manipulate the free parameters of the working system in order to approximately reach the Carnot limit---see Fig. \ref{FIG2}. Unlike in the case of one-qubit system, since $f_2(x)$ is always positive, one can thus tune the free parameters of our system---$J_{1}$ and $J_{2}$---such that the second order correction $\mathbbmss{S}_{2}(x)$ becomes negligible and the deviation from the Carnot efficiency vanishes up to second order in the perturbation parameter $\lambda$. Note that such an efficiency can be achieved without asking that $J_{1,2}=0$, rather by only requiring a degree of parameter controllability able to make Eq. \eqref{2qpertent2} vanish.
 
\section{Summary}

We have proposed a model of quantum heat engine operating an optimal cycle, reaching Carnot efficiency, with respect to a given driving Hamiltonian, and have computed the deviations from the maximal efficiency in the presence of a generic perturbation. We have shown that, unlike in the case of one-qubit working substance, when the working substance is a pair of noninteracting qubits, by suitably adjusting the free parameters of a Heisenberg-like interaction, one can suppress the deviation from the Carnot efficiency up to third-order perturbation.

\textit{Acknowledgements.}---M.R. acknowledges helpful discussions with M. S. Salehi Kadijani. This work was partially supported by Iran's National Elites Foundation (to M.R.), the Project EU 2020 ERC-2015-STG G.A. No. 677488 (to S.M.), and Sharif University of Technology's Office of Vice President for Research and Technology through Grant No. QA960512 (to A.T.R.).


\appendix
\begin{widetext}

\section{Summary of equilibrium thermodynamic relations} 
\label{sec:H-EX}

Given a quantum system described by a density matrix $\varrho$ and a Hamiltonian operator $H$, from the differential form of the first law of thermodynamics 
\begin{equation}
\label{1law}
\mathrm{d}\mathbbmss{U}=\dbar \mathbbmss{Q}+\dbar \mathbbmss{W}, 
\end{equation}
where the internal energy is $\mathbbmss{U}=\mathrm{Tr}[\varrho H]$, the infinitesimal amounts of exchanged heat $\dbar \mathbbmss{Q}$ and work $\dbar \mathbbmss{W}$, respectively, read as
\begin{align}
\label{1law1}
\dbar \mathbbmss{Q}&=\mathrm{Tr}[\mathrm{d}\varrho \,H],\\
\dbar \mathbbmss{W}&=\mathrm{Tr}[\varrho\, \mathrm{d}H].
\end{align}
If the system is in a Gibbs thermal state at inverse temperature $\beta=1/T$, i.e., $\varrho=(1/Z)\,\mathrm{e}^{-\beta H}$ (with $Z=\mathrm{Tr}[\mathrm{e}^{-\beta H}]$), one can readily see
\begin{equation}
\label{1law2}
\dbar \mathbbmss{Q}=-(1/\beta)\mathrm{Tr}[\mathrm{d}\varrho\, \log\varrho] = (1/\beta)\, \mathrm{d}\mathbbmss{S},
\end{equation}
where $\mathbbmss{S}=-\mathrm{Tr}[\varrho\log\varrho]$ is the von Neumann entropy of $\varrho$. From the above expression and Eq. \eqref{1law}, one obtains the differential free energy 
\begin{equation}
\label{1law3}
\mathrm{d}\mathbbmss{F}=\mathrm{d}\big(\mathbbmss{U}-(1/\beta) \mathbbmss{S}\big)=-\mathbbmss{S}\,\mathrm{d}T+\dbar \mathbbmss{W}.
\end{equation}
This implies that in an isothermal quasi-static process (where $\mathrm{d}\beta=0$), we have
\begin{equation}
\label{1law4}
\mathrm{d} \mathbbmss{F} =\dbar \mathbbmss{W}=-(1/\beta)\,\mathrm{d}\log Z.
\end{equation}

\section{Dyson expansion of thermodynamic quantities}
\label{sec:QHE}

Given the perturbed Hamiltonian $H^{[\lambda]}=g H+\lambda V$, the Dyson expansion in a series of powers of $\lambda$ is obtained as follows.
First, note that we can write
\begin{eqnarray*}
\mathrm{e}^{-\beta  H^{[\lambda]}}-\mathrm{e}^{-\beta g H}&=& \int_0^{1}\mathrm{d}s \frac{\mathrm{d}}{\mathrm{d}s}\bigg[\mathrm{e}^{-(1-s)\beta g H} \mathrm{e}^{-s\beta H^{[\lambda]}}\bigg] =\beta\int_0^{1}\mathrm{d}s\,\mathrm{e}^{-(1-s)\beta g H}(g H-H^{[\lambda]})\,\mathrm{e}^{-s\beta H^{[\lambda]}}\\
&=-&\lambda \beta \int_0^{1}\mathrm{d}s\,\mathrm{e}^{-(1-s)\beta g H} V \mathrm{e}^{-s\beta H^{[\lambda]}}.
\end{eqnarray*}
Iterating the procedure yields the Dyson series, $\mathrm{e}^{-\beta H^{[\lambda]}}=\sum_{\ell=0}(\lambda\beta)^{\ell}\,E_\ell(\beta g)$,
where
\begin{equation}
\label{Dyson}
E_{\ell} (\beta g)=(-1)^{\ell}\mathrm{e}^{-\beta g H} \int_0^1\mathrm{d}s_1\int_0^{s_1}\mathrm{d}s_2\cdots\int_0^{s_{\ell-1}}\mathrm{d}s_{\ell}\,V(s_1)V(s_2)\cdots V(s_{\ell}),
\end{equation}
where $V(s)=\mathrm{e}^{s\beta g H} V \mathrm{e}^{-s\beta gH}$. It then follows that the partition function can be expanded as 
\begin{equation}
\label{partition1}	
Z^{[\lambda]} =\mathrm{Tr}[\mathrm{e}^{-\beta H^{[\lambda]}}] =\textstyle{\sum_{\ell=0}}(\lambda\beta)^{\ell}Z_\ell(\beta g),
\end{equation}
with
\begin{equation}
Z_\ell(\beta g)=(-1)^\ell\mathrm{Tr}\left[\mathrm{e}^{-\beta g H} \int_0^1\mathrm{d}s_1\int_0^{s_1}\mathrm{d}s_2\cdots\int_0^{s_{\ell-1}}\mathrm{d}s_{\ell}\,V(s_1)\,V(s_2)\cdots V(s_{\ell})\right].
\end{equation}
For convenience, we report the first three contributions to the perturbed partition function,
\begin{align}
\label{pertpart0}
Z_0(\beta g)=&\mathrm{Tr}[\mathrm{e}^{-\beta g H}],
\\
\label{pertpart1}
Z_1(\beta g)=&-\,\mathrm{Tr}[\mathrm{e}^{-\beta g H} V\Big],
\\
\label{pertpart2}
Z_2(\beta g)=&\mathrm{Tr}\Big[\mathrm{e}^{-\beta g H} \int_0^1\mathrm{d}s_1\int_0^{s_1}\mathrm{d}s_2\, V(s_2)\,V\,\Big].
\end{align}
By means of the series expansion of the perturbed partition function, one also expands the von Neumann entropy $\mathbbmss{S}^{[\lambda]}$ of the perturbed state by using that the differential of the free energy $\mathbbmss{F}=\mathbbmss{U}-(1/\beta)\mathbbmss{S}=-(1/\beta)\log Z$ yields $\displaystyle \mathbbmss{S}=\beta^2\partial_\beta \mathbbmss{F}$. Keeping the perturbation expansion of the partition function up to the first nonvanishing order, say the $k$th,  
\begin{equation}
Z^{[\lambda]}\approx  Z_0+(\lambda\beta)^{k} Z_k,
\end{equation}
the free energy is obtained as
\begin{equation}
\mathbbmss{F}^{[\lambda]}\approx  -\frac{1}{\beta}\left(\log Z_0(\beta g)+\frac{Z_k(\beta g)}{Z_0(\beta g)}\right),
\end{equation}
whence
\begin{equation}
\mathbbmss{S}^{[\lambda]}\approx  \underbrace{\log Z_0(\beta g)-\beta\,\partial_\beta\log Z_0(\beta g)}_{\mathbbmss{S}_0(\beta g)} + (\lambda\beta)^k\underbrace{\left((1-k)\frac{Z_k(\beta g)}{Z_0(\beta g)}-\beta\partial_\beta\left(\frac{Z_k(\beta g)}{Z_0(\beta g)}\right)\right)}_{\mathbbmss{S}_k(\beta g)}.
\end{equation}
Note that the coefficients of the expansion are functions of the product $x=\beta g$,
\begin{align}
\label{appBent}
\mathbbmss{S}_0(x)&=\log Z_0(x)-x\,\partial_x\log Z_0(x),\\
\mathbbmss{S}_k(x)&=(1-k)\frac{Z_k(x)}{Z_0(x)}-x\,\partial_x\left(\frac{Z_k(x)}{Z_0(x)}\right).
\end{align}

\twocolumngrid
\end{widetext}


\begin{thebibliography}{100}

\bibitem{1959-Scovil} H. E. D. Scovil and E. O. Schulz-DuBois, Phys. Rev. Lett. \textbf{2}, 262 (1959).

\bibitem{1979-Alicki} R. Alicki, J. Phys. A: Math. Gen. \textbf{12}, L103 (1979).

\bibitem{1984-Kosloff} R. Kosloff, J. Chem. Phys. \textbf{80}, 1625 (1984).

\bibitem{2016-Alipour} S. Alipour, F. Benatti, F. Bakhshinezhad, M. Afsary, S. Marcantoni, and A. T. Rezakhani, Sci. Rep. \textbf{6}, 35568 (2016).


\bibitem{2016-Rosnagel} J. Ro{\ss}nagel, S. T. Dawkins, K. N. Tolazzi, O. Abah, E. Lutz, F. Schmidt-Kaler, and K. Singer, Science \textbf{352}, 325 (2016).

\bibitem{2000-Bender} C. M. Bender, D. C. Brody, and B. K. Meister, J. Phys. A: Math. Gen. \textbf{33}, 4427 (2000).

\bibitem{2006-Quan} H. T. Quan, Y. D. Wang, Y.-X. Liu, C. P. Sun, and F. Nori, Phys. Rev. Lett. \textbf{97}, 180402 (2006).

\bibitem{2007-Quan} H. T. Quan, Y.-X. Liu, C. P. Sun, and F. Nori, Phys. Rev. E \textbf{76}, 031105 (2007).

\bibitem{2009-Quan} H. T. Quan, Phys. Rev. E \textbf{79}, 041129 (2009).

\bibitem{2012-Brunner} N. Brunner, N. Linden, S. Popescu, and P. Skrzypczyk, Phys. Rev. E \textbf{85}, 051117 (2012).

\bibitem{2013-Huang} X. L. Huang, L. C. Wang, and X. X. Yi, Phys. Rev. E \textbf{87}, 012144 (2013).

\bibitem{2015-Altintas} F. Altintas and \"O. E. M\"ustecaplioglu, Phys. Rev. E \textbf{92}, 022142 (2015).

\bibitem{1992-Kosloff} E. Geva and R. Kosloff, J. Chem. Phys. \textbf{96}, 3054 (1992).

\bibitem{1996-Kosloff} T. Feldmann, E. Geva, and R. Kosloff,  Am. J. Phys. \textbf{64}, 485 (1996).

\bibitem{Feldmann} T. Feldmann and R. Kosloff, Phys. Rev. E \textbf{61}, 4774 (2000); 
\textit{ibid.} \textbf{68}, 016101 (2003).

\bibitem{2011-Skrzypczyk} P. Skrzypczyk, N. Brunner, N. Linden, and S. Popescu, J. Phys. A: Math. Theor. \textbf{44}, 492002 (2011).

\bibitem{2011-Benenti} G. Benenti, K. Saito, and G. Casati, Phys. Rev. Lett. \textbf{106}, 230602 (2011).

\bibitem{2012-Wang} J. Wang, Z. Wu, and J. He, Phys. Rev. E \textbf{85}, 041148 (2012). 

\bibitem{2013-Gelbwaser} D. Gelbwaser-Klimovsky, R. Alicki, and G. Kurizki, Phys. Rev. E \textbf{87}, 012140 (2013).

\bibitem{2013-Allahverdyan} A. E. Allahverdyan, K. V. Hovhannisyan, A. V. Melkikh, and S. G. Gevorkian, Phys. Rev. Lett. \textbf{111}, 050601 (2013).

\bibitem{2014-Kosloff} R. Kosloff and A. Levy, Annu. Rev. Phys. Chem. \textbf{65}, 365 (2014).

\bibitem{2015a-Proesmans} K. Proesmans and C. Van den Broeck, Phys. Rev. Lett. \textbf{115}, 090601 (2015).

\bibitem{2015b-Polettini} M. Polettini, G. Verley, and M. Esposito, Phys. Rev. Lett. \textbf{114}, 050601 (2015).

\bibitem{2015c-Brandner} K. Brandner, K. Saito, and U. Seifert,  Phys. Rev. X \textbf{5}, 031019 (2015).

\bibitem{2016-Campisi} M. Campisi and R. Fazio,  Nature Commun. \textbf{7}, 11895 (2016).

\bibitem{2012-Abe} S. Abe and S. Okuyama, Phys. Rev. E \textbf{85}, 011104 (2012).

\bibitem{2013-Goswami} H. P. Goswami and U. Harbola, Phys. Rev. A \textbf{88}, 013842 (2013).

\bibitem{2011-DeLiberato} S. De Liberato and M. Ueda, Phys. Rev. E \textbf{84}, 051122 (2011).

\bibitem{2012-Zagoskin} A. M. Zagoskin, S. Savel'ev, F. Nori, and F. V. Kusmartsev, Phys. Rev. B \textbf{86}, 014501 (2012).

\bibitem{2014-Rosnagel} J. Ro{\ss}nagel, O. Abah, F. Schmidt-Kaler, K. Singer, and E. Lutz, Phys. Rev. Lett. \textbf{112}, 030602 (2014).

\bibitem{2015-Uzdin} R. Uzdin, A. Levy, and R. Kosloff, Phys. Rev. X \textbf{5}, 031044 (2015).

\bibitem{2017-Watanabe} G. Watanabe, B. P. Venkatesh, P. Talkner, and A. del Campo, Phys. Rev. Lett. \textbf{118}, 050601 (2017).

\bibitem{Callen} H. B. Callen, \emph{Thermodynamics and an Introduction to Thermostatistics} (John Wiley, New York, 1985).

\end{thebibliography}
\end{document}